\documentclass
[preprintnumbers,amsfonts,amssymb,superscriptaddress,twocolumn,10pt,tightenlines,nobibnotes,floats,final,prb,superbib]{revtex4}%
\usepackage{amsfonts}
\usepackage{amsmath}
\usepackage{amssymb}
\usepackage{graphicx}
\usepackage{dcolumn}
\usepackage{bibmods}%
\begin{document}
\preprint{ }
\title[FIR modes in 2DEHGS]{Magneto-infrared modes in InAs-AlSb-GaSb coupled quantum wells.}
\author{L.-C. Tung}
\affiliation{National High Magnetic Laboratory, Tallahassee, Florida 32310}
\author{P. A. Folkes}
\affiliation{U.S. Army Research Laboratory, Adelphi, Maryland 20783}
\author{G. Gumbs}
\affiliation{Hunter College, City University of New York, New York, New York 10021}
\author{W. Xu}
\affiliation{Institute of Solid State Physics, Chinese Academy of Sciences, Hefei and
Yunnan University, Kunming, China}
\author{Y.-J. Wang}
\affiliation{National High Magnetic Laboratory, Tallahassee, Florida 32310}
\date{\today}

\begin{abstract}
We have studied a series of InAs/GaSb coupled quantum wells using
magneto-infrared spectroscopy for high magnetic fields up to $33%
\operatorname{T}%
$ within temperatures ranging from $4%
\operatorname{K}%
$ to $45%
\operatorname{K}%
$ in both Faraday and tilted field geometries. This type of coupled quantum
wells consists of an electron layer in the InAs quantum well and a hole layer
in the GaSb quantum well, forming the so-called two dimensional electron-hole
bilayer system. Unlike the samples studied in the past, the hybridization of
the electron and hole subbands in our samples is largely reduced by having
narrower wells and an AlSb barrier layer interposed between the InAs and the
GaSb quantum wells, rendering them weakly hybridized. Previous studies have
revealed multiple absorption modes near the electron cyclotron resonance of
the InAs layer in moderately and strongly hybridized samples, while only a
single absorption mode was observed in the weakly hybridized samples. We have
observed a pair of absorption modes occurring only at magnetic fields higher
than $14%
\operatorname{T}%
$, which exhibited several interesting phenomena. Among which we found two
unique types of behavior that distinguishes this work from the ones reported
in the literature. This pair of modes is very robust against rising thermal
excitations and increasing magnetic fields alligned parallel to the
heterostructures. While the previous results were aptly explained by the
antilevel crossing gap due to the hybridization of the electron and hole
wavefunctions, i.e. conduction-valence Landau level mixing, the unique
features reported in this paper cannot be explained within the same concept.
The unusual properties found in this study and their connection to the known
models for InAs/GaSb heterostructures will be disccused; in addition, several
alternative ideas will be proposed in this paper and it appears that a
spontaneous phase separation can account for most of the observed features.

\end{abstract}

\volumeyear{year}
\volumenumber{number}
\issuenumber{number}
\eid{identifier}
\received[Received text]{date}

\revised[Revised text]{date}

\accepted[Accepted text]{date}

\published[Published text]{date}

\startpage{101}
\endpage{ }
\maketitle

\section{Introduction}

Bose-Einstein condensate (BEC) is a highly ordered state in which the
wavefunction phase is coherent over distances much longer than the separation
between individual particles. BEC of excitons in semiconductors has been a
subject of intense interest for decades. In semiconductors, an electron in
conduction band and a hole in valence band can bind together by Coulomb
interaction to form an exciton, a composite boson in contrast to electrons in
superconductors, which bind together to form Cooper pairs.\cite{Bar57} While
BEC and superconductivity were found in superconductors, excitons in
semiconductors have the potential to reach BEC, novel collective ground states
or even superfluidity. This concept has been discussed for three-dimensional
bulk
materials,\cite{Bla62,Kel65,Koz65,Kel68,Jer67,Fen68,Nak76,Vol76,Kor93,She94,Kor94,Bal00}
and also for two-dimensional bilayer
heterostructures.\cite{Loz75,Yos78,Ler79,Bas82,Byc83,Dat85,Paq85,Zhu90,Yos90,Bau90,Che91,Xia92,Nav94,Nav96,Zhu95,Chu96,Leo00,Pal02,Bal04,Eis04,Lai05,Loz07}
Several novel phenomena or collective states have been proposed, including
exciton insulator,\cite{Kel65,Koz65,Kel68,Jer67,Fen68,Vol76,Zhu90} gas-liquid
type phase transition,\cite{Nak76} BEC of
magnetoexcitons,\cite{Bla62,Ler79,Paq85,Yos90,Zhu95,Chu96,Pal02,Eis04,Lai05}
electric dipole-density-wave,\cite{Yos78} electron-hole
magnetoplasma,\cite{Bau90} double charge-density-wave state\cite{Che91} and
most interestingly superfluidity.\cite{Kor93,Kor94,Bal04,Loz07}

Signatures possibly resulting from the Bose-Einstein (BE) statistics of
excitons have been observed in several bulk systems, such as Ge, CuCl,
CuO$_{2}$ and in a form of ferromagnetism in La$_{x}$Ca$_{1-x}$B$_{6}$ and
La$_{x}$Sr$_{x}$B$_{6}$.\cite{Tim83,Pey83,Sno87,For93,You99} Excitons are
usually created by shining light on the semiconductor, which creates excess
electrons and holes in equal numbers. Optically generated excitons are
ephemeral and decay quickly via emission of light. Rapid recombination of
electrons and holes tends to destroy the coherence of the excitonic
states.\cite{Xia92} The lifetime of the excitons in 3D semiconductors is
usually shorter than thermalization and condensate time, making it difficult
to investigate BEC or BE statistics in 3D systems. On the other hand, the
lifetime of excitons in 2D heterostructures can be enhanced if the rate of
recombination is reduced by spatially separating electrons and
holes.\cite{Fuk90} Electrons and holes can be confined in different layers
between which tunneling can be made negligible by inserting a
barrier.\cite{Siv92,Kan94} Though this process reduces Coulomb interaction,
and thus a lower critical temperature $T_{C}$, confinement of both electrons
and holes increases exciton binding energy compared to its value in 3D bulk
systems.\cite{Xia92}

It is by far, most favorable to form an electron-hole droplet in a homogeneous
bulk semiconductor, which is an insulator;\cite{Kel65,Koz65,Kel68,Jer67,Fen68}
however, superconductivity or superfluidity are in principle possible in
bilayer heterostructures.\cite{Loz75,Paq85,Nav96,Bal04} The spatial separation
causes the excitons to act like oriented electric dipoles, which have
repulsive interactions between each other, preventing the electrons and holes
from agglomerating into a universal electron-hole plasma.\cite{Eis04} It has
been suggested that the interaction leading to the formation of the droplet
can be suppressed by a strong magnetic field\cite{Kor93,Kor94} or
sophisticated heterostructures.\cite{Zhu95,Loz07} However, one should note
that it is generally believed that an ideal BEC is not allowed in a true 2D
system, so other interesting phenomena such as Berezinskii-Kosterlitz-Thouless
transition can also be expected.\cite{Ber71,Kos73}

With the advantages over a 3D bulk system and via the modern crystal growth
techniques\cite{Fuk90,Siv92} such as molecular beam epitaxy, coupled quantum
well (CQW) heterostructures have emerged a promising candidate for achieving
exciton BEC in
semiconductors.\cite{Fuk90,Gol90,But01,But02,But04,Sno02,Yan06,But07,Ste08}
The CQW consists of two low-band-gap well layers separated by a high band gap
barrier layer. Photoexcited indirect excitons are formed between the holes in
the valence band of one layer and the electrons in the conduction band of the
other. These excitons carry electric dipoles which are aligned perpendicular
to the QW plane, stabilizing excitonic states against the formation of
droplets,\cite{Yos90,Zhu95} reinforcing the BEC\cite{Leg01} and resulting in
the screening of an in-plane potential.\cite{But02} Recent discoveries in CQW
heterostructures have revealed several interesting phenomena leading to an
exciton BEC state in
semiconductors.\cite{Fuk90,Gol90,But01,But02,But04,Sno02,Yan06,But07,Ste08}
Others have found exciton-polariton BEC by using a microcavity to couple heavy
hole excitons to cavity
photons.\cite{Den02,Lai07,Bal07,Den07,Lag08,Uts08,Amo09} A list of several BEC
systems and their properties was reviewed by Keeling and Berloff recently.
\cite{Kee09}

Among the candidates possibly reaching exciton BEC in 2D systems, InAs/GaSb
CQW has a unique type-II band alignment, in which the bottom of the conduction
band in bulk InAs lies below the top of the valence band in bulk GaSb. (i.e.
negative effective band gap $E_{g}\sim-150$ to $-180$m$%
\operatorname{eV}%
$)\cite{Alt83,Nil06} The overlap of electron and hole band edges can
spontaneously form excitons without photoexcitations if the single particle
energy gap is smaller than the exciton binding energy. These spontaneously
formed excitons have no recombination, and hence, possess long lifetimes. As
electrons are transferred to the empty conduction band of the InAs layer
leaving behind holes in the GaSb layer, electron and hole densities in the
order of $\sim10^{11}%
\operatorname{cm}%
^{-2}$ can be obtained\cite{Fol06} without intensional doping or applying a
gate voltage. This spatially separated two dimensional electron-hole system
(2DEHS) is confined within different layers, and the 2DES and 2DHS are in
equilibrium with each other at the interface. The energy gap between electron
and hole subbands can then be tuned by varying the InAs/GaSb well width, or by
doping GaSb layers with Al to achieve semiconductor or semimetallic structure.
(effective band gap $E_{g}=-150$ to $300$m$%
\operatorname{eV}%
)$\cite{Kon94,Che95,Kon97} This type of CQW system has several advantages. It
is more desirable to produce exciton fluids at high densities, since the
energy of the condensate would be larger, i.e. larger critical temperature,
while optically pumped indirect exciton systems usually have lower exciton
density. ($\sim10^{10}%
\operatorname{cm}%
^{-2}$)\cite{Fuk90,But04} The variation of the key properties with exciton
densities can be systematically investigated by tuning the bias voltage with
appropriate gates or the thickness of the GaSb cap layer,\cite{Fol06} while
high-power photoexcitation tends to drive the optically pumped exciton fluids
into electron-hole plasma regime.\cite{Ste08} In addition, InAs/GsSb systems
have potential applications for intersubband tunnel diodes\cite{Kit97} and
mid-infrared optical devices.\cite{Ohn92,Hal00}

However, its advantages are also disadvantages. Excitons formed in equilibrium
lead to difficulties in observation, since they are dark excitons which do not
luminesce. Fewer means are available for investigating the properties of these
dark excitons. Though several types of transport property measurements have
been proposed and carried out for exciton CQW,\cite{Eis04,Spi00,Kel04,Tut04}
the CQW samples were primarily investigated by optical means. One of the
optical techniques for investigating these dark excitons is the cyclotron
resonance (CR) spectroscopy by measuring magneto-infrared (IR) absorption.
However, while spatially resolved photoluminescence techniques are used to
investigate the macroscopic coherence of the photoexcited excitons, the
spatially resolved IR technique has only been developed in the mid-infrared
range with a spatial resolution as good as microns.\cite{Tal00,Sam06}

In this type of 2DEHS, several far-infrared (FIR) active modes around CR have
been observed and extensively studied in the past two decades. On one hand,
the results suggest the formation of stable excitons by showing exciton's
$1s-2p$ internal transitions;\cite{Kon94,Che95,Kon97} on the other hand, they
were interpreted as the hybridization gap due to a mixing of the electron and
hole
wavefunctions.\cite{Cla86,Sun92,Chi96,Yan97,Lak97,Mar99,Vas99,Pou99,Pet02,Suz03,Pet04,Pet07}
Though most of the FIR investigation reveal multiple modes around CR in
InAs/GaSb CQWs, only a single mode was observed up to $12%
\operatorname{T}%
$ and interpreted as electron-CR by Heitmann \textit{et al}.\cite{Hei86}
Earlier studies in InAs/GaSb superlattice systems show electron CR,
intersubband transitions and sometimes two absorptions near CR attributed to
two occupied electron subbands.\cite{Blu79,Gul80,Maa81,Maa82,Blu82} More
recent reports about the InAs/GaSb 2DEHS demonstrate results in favor of the
hybridization gap interpretation.\cite{Pet04,Pet07} The hybridization model in
association with the $\mathbf{k\cdot p}$ model can account for several
features observed in this type of material, including the magnetic-field
dependence of the CR energies,\cite{Chi96,Pet04} oscillation in CR linewidth,
mass, and amplitude,\cite{Hei86,Kon94,Che95,Kon97,Vas99} the influence of a
parallel magnetic field,\cite{Lak97,Mar99,Pou99,Pet07} the slow decay rate of
the temperature-dependent absorption intensity,\cite{Mar99,Pet07} and the
anomalies observed in magneto-transport
measurements.\cite{Men85,Dal96,Nic00,Tak03}

In InAs/GaSb CQW systems, the overlap between the electron and hole subbands
can be reduced or even completely eliminated by the confinement energy (small
well width and high magnetic field); therefore minimizing the effect of the
hybridization. It has been shown that the degree of hybridization can be
modified by the width of InAs and GaSb well, or by inserting a spacer
layer.\cite{Mar99,Vas99,Nil06,Suz03,Pet04,Pet07} Increasing InAs or GaSb well
width tends to enhance the degree of hybridization by increasing the mixing of
the electron and hole Landau levels, whereas the interposed barrier layer
tends to reduce it. With a barrier layer as thin as $2$-$4$ atomic layers, the
mixing of the electron and hole wavefunctions can be suppressed,\cite{Nav96}
since the penetration length should be of the order of $1%
\operatorname{nm}%
$ in this system.\cite{Vas99} Although the electron-hole Coulomb interaction
decreases with increasing average distance between 2DES and 2DHS as the
barrier layer is inserted, such a thin barrier layer has a negligible effect
on the interlayer Coulomb interactions.\cite{Nav96} Generally, multiple
absorption modes near CR were found in systems, which have configurations in
favor of hybridization model in contrast to a single absorption mode,
attributed to electron CR, which was observed in the weakly hybridized
systems.\cite{Mar99,Vas99,Pet04}

In this paper, we present the result of the FIR magneto-optical study on
weakly hybridized InAs/AlSb/GaSb CQWs, which have narrow well width and a $1%
\operatorname{nm}%
$ AlSb barrier layer separating electron and hole layers. This particular
configuration has been shown theoretically to have negligible hybridization
coupling, but sufficient interlayer Coulomb coupling, proposed as a candidate
for achieving a stable condensed phase of excitons.\cite{Nav96} We have
observed a pair of FIR absorptions occurring only at magnetic fields higher
than $14%
\operatorname{T}%
$ with a field-independent (piecewise) energy separation. This pair of modes
does not decay with increasing temperature up to $45%
\operatorname{K}%
$ and is insensitive to increasing parallel magnetic fields. These features
distinguish our results from the ones reported in the literature. The
connection between these phenomena and several models known to cause multiple
absorption modes near CR is discussed. None of the models can account for the
key features observed in this work, which suggests that more investigations
should be carried out for this exciton system.

\section{Experimental}

The active region of the InAs-AlSb-GaSb CQW consists a $170%
\operatorname{\text{\AA}}%
$ InAs quantum well (electron layer) and a $50%
\operatorname{\text{\AA}}%
$ GaSb quantum well (hole layer), separated by a $10%
\operatorname{\text{\AA}}%
$ AlSb barrier, and then completed by AlSb layers as outer barriers. The
heterostructure's band diagram is shown in Fig.1. The electron density is
around $8.6\times10^{11}%
\operatorname{cm}%
^{-2}$ and the hole density is around $1.2\times10^{11}%
\operatorname{cm}%
^{-2}$ determined by the transport measurement\cite{Fol06} for the particular
sample shown in this paper. The electron and hole densities can be determined
separately from the SdH oscillations. This implies a very weak hybridization
effect between the 2DES and 2DHS states. Samples whose electron densities
ranging from $8\times10^{11}%
\operatorname{cm}%
^{-2}$ to $9.5\times10^{11}%
\operatorname{cm}%
^{-2}$ and hole densities ranging from $1.2\times10^{11}%
\operatorname{cm}%
^{-2}$ to $0.2\times10^{11}%
\operatorname{cm}%
^{-2}$ have been investigated. The carrier densities are controlled by the
thickness of the GaSb cap layer and determined from the transport measurements
fitted by two carrier conduction model.\cite{Zim72} The detail configuration
of the heterostructure and its corresponding carrier densities can be found in
ref. [65]. The electron mobility is around $10^{5}%
\operatorname{cm}%
^{2}/%
\operatorname{V}%
\operatorname{s}%
$.\cite{Fol06} The FIR transmission spectroscopy is carried out by a
commercial Fourier transform interferometer up to $33%
\operatorname{T}%
$ from $4%
\operatorname{K}%
$ to $45%
\operatorname{K}%
$ using light-pipe optics and a Si bolometer. The magneto-FIR modes are
extracted by the ratio of the spectra measured with and without a magnetic field.

\begin{figure}[tb]
{{{\includegraphics[
natheight=6.12in,
natwidth=7.96in,
height=2.2in,
width=3.4in
]{C: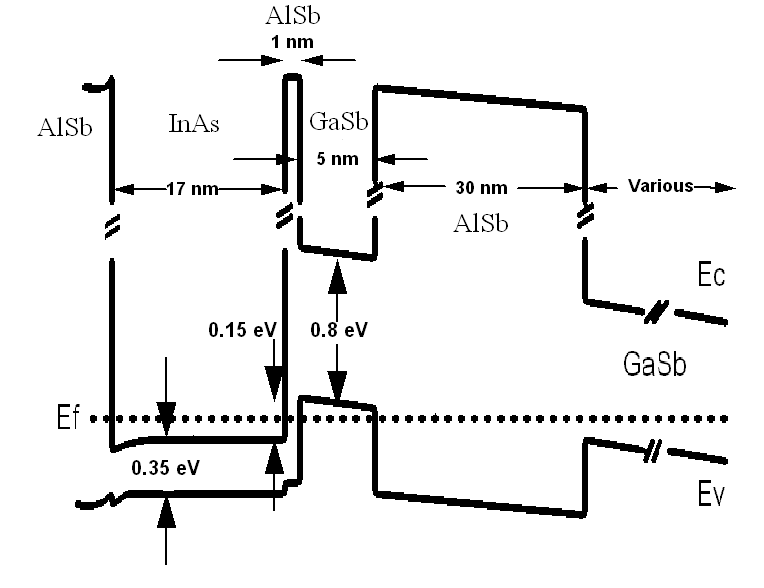}}}}\caption{Band diagram for the active region of
the InAs-AlSb-GaSb heterostructure. $E_{C}$ and $E_{V}$ are the respective
conduction band minimum and top of the valence band in the GaSb cap layer.
$E_{f}$ marks the position of the Fermi level.}%
\end{figure}

\section{Results and Discussion}

A typical set of the magneto-spectra measured within the range of values
$14\leq B\leq28%
\operatorname{T}%
$ is displayed in Fig. 2 and a pair of FIR active modes near the CR energy can
be observed in the Faraday geometry. Below $10%
\operatorname{T}%
$, only a single absorption mode is observed, attributed to the electron CR in
the InAs QW, as shown in the inset of Fig. 1. From the CR energies at low
magnetic fields, the effective mass is determined to be $0.037m_{e}$,
consistent with the values reported in the literature, but much larger than
the bulk electron mass $\sim0.02m_{e}$. The observed absorption modes below
and above the GaAs restrahlen band exhibits an oscillation in CR amplitude,
linewidth and masses, which is similar to the one reported in the
literature.\cite{Kon94,Che95,Kon97,Hei86} All of these spectra are dark, which
were taken without the beforehand light emitting diode (LED) illumination.
Unlike some reported results,\cite{Kon94,Che95,Kon97,Vas99} beforehand LED
illumination does not induce new modes, nor does it change the CR energy or
absorption intensity. Several works reported the observation of the hole CR in
the GaSb layer and the hole effective mass was determined to be around $0.1$
to $0.2m_{e}$.\cite{Che95,Sun92} No signs of hole CR have been observed beyond
the noise level in any of the samples investigated, though high magnetic
fields used in these measurements can place the hole CR in the observable
frequency range in our setups. \begin{figure}[tb]
{{{\includegraphics[
natheight=11in,
natwidth=8.5in,
height=4in,
width=3.4in
]{C: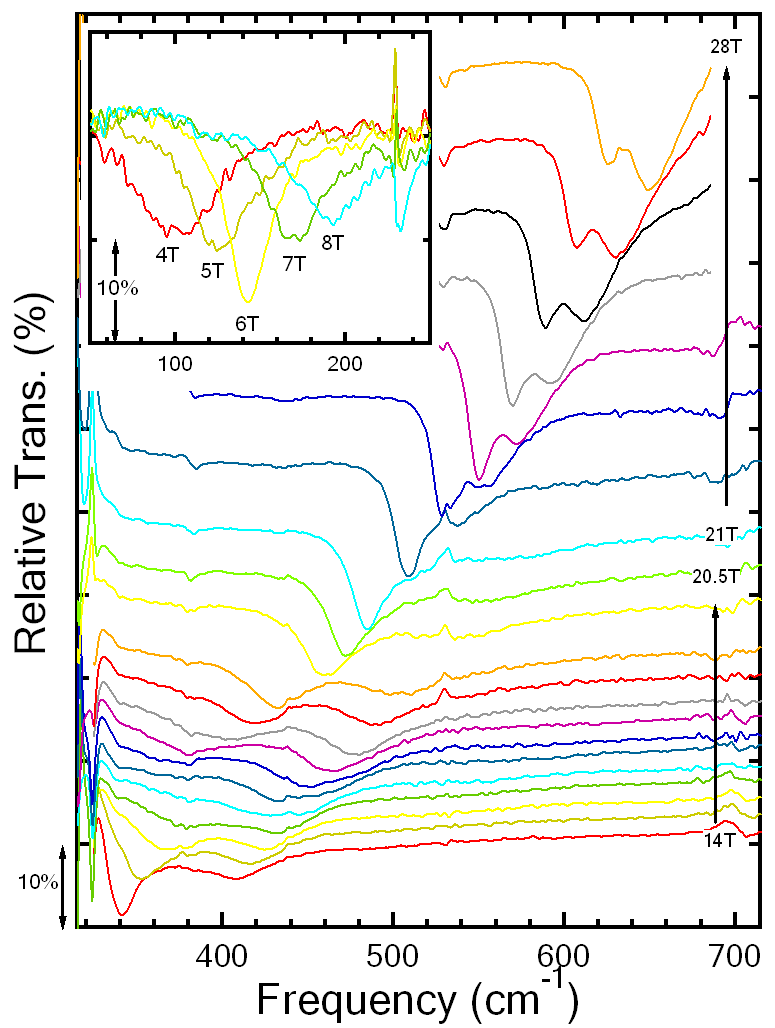}}}}\caption{Magneto-FIR spectra from
$14\operatorname{T}$ to $28\operatorname{T}$ are displayed. The speactra are
vertically shifted to demonstrate the evolution of the absorption modes with
increasing magnetic field. Inset: Spectra measured at low magnetic fields
($4\operatorname{T}$ to $8\operatorname{T}$) are displayed and only a single
absorption mode can be observed.}%
\end{figure}

This result is similar to the general behavior of the reported multiple
absorption modes. A pair of modes (in some reports, more than two modes),
separated by a few m$%
\operatorname{eV}%
$, occurs near CR energy and their relative strength evolves with an
increasing magnetic field. In this work, the lower-energy transition (CRL) is
the strongest of the two at $14%
\operatorname{T}%
$. With increasing magnetic field, the higher-energy transition (CRH)
increases in intensity at the expense of \ CRL. This process is reversed at
around $17%
\operatorname{T}%
$, above which CRL increases in intensity at the expense of CRH. The process
is then reversed again at around $23%
\operatorname{T}%
$ and CRH remains the dominating transition up to $33%
\operatorname{T}%
$, which is the highest field measured for this series of samples. The
evolution of the relative intensity is more similar to previous works, which
interpreted the pair of modes in terms of CR and exciton's internal
transitions;\cite{Kon94,Che95,Kon97} while others, which interpreted the
multiple absorption modes in terms of hybridization of electron and hole
wavefunctions, reported an evolution that a new mode appears from the
lower-energy side, and replaces those that tail off to the higher-energy
region.\cite{Pet04}

Although this pair of modes exhibit features similar to the ones reported in
the literature, several key differences are difficult to explain by the
previous models, hybridization of the electron-hole wavefunctions or exciton's
internal transitions. These key differences will be discussed in the following subsections.

\subsection{The occurrence of the CR splitting}

One may notice that we prefer to refer the pair of modes found in this work as
"CR splitting", in place of the term, "multiple absorption modes near CR",
since we believe that they are both a consequence of electron CR involved with
a transition between two electron Landau levels. (LLs) In the hybridization
model, the absorption modes arose from transitions between two electron-like
LLs and between a hole-like and an electron-like LL; while in the exciton
model, one of the modes is due to the electron CR and the other, from the
$1s-2p$ internal transitions of an exciton. For those transitions, we will
refer to them as multiple absorption modes near CR, though we will also use
this term to describe our findings occasionally, since it is a more general
term. In addition, we will also refer to the two models in abbreviated terms
as "exciton model" and "hybridization model".

It has been argued that hybridization model, or conduction-valence Landau
level (LL) mixing, has an insignificant effect at high and low
fields,\cite{Chi96} while the CR splitting in this paper occur only at
magnetic fields higher than $14%
\operatorname{T}%
$. In the literature, multiple absorptions were observed throughout and within
nearly entire magnetic range
investigated.\cite{Kon94,Che95,Kon97,Cla86,Sun92,Chi96,Yan97,Lak97,Mar99,Vas99,Pou99,Pet02,Suz03,Pet04,Pet07}
One may argue that the magnetic field range, where the CR splitting were
observed in this paper, is in fact the "intermediate" field range, but it is
very unlikely. It is essentially impossible to place an intermediate regime
within the confine of the given system.

The hybridization model is about an antilevel crossing gap which occurs due to
a resonance between electron and hole wavefunctions. It occurs at a magnetic
field when the electron or hole LLs are aligned; in other words, at the points
where the electron and hole LLs cross each other. Using electron mass
$m_{e}^{\ast}=0.037m_{e}$, hole mass $m_{h}^{\ast}=0.1m_{e}$ and electron
density as $8.6\times10^{11}%
\operatorname{cm}%
^{-2}$, an energy diagram for the Landau levels in this system is established
in Fig. 3. Near the crossing points, a transition between a hole-like to an
electron-like LL becomes possible due the mixing of the electron and hole LL
wavefunctions. As a result, more absorption modes make their appearance other
than the transitions via two electron LLs. For the exciton model, the binding
of electrons and holes requires the energy separation between them to be lower
than the binding energy. We can expect anomalies as a result of electron-hole
binding around these crossing points as well, though there should be a sizable
excitonic binding energy before the alignment of the electron and hole
LLs.\cite{Xia92}

\begin{figure}[tb]
{{{\includegraphics[
natheight=8.5in,
natwidth=11in,
height=2.473in,
width=3.4in
]{C: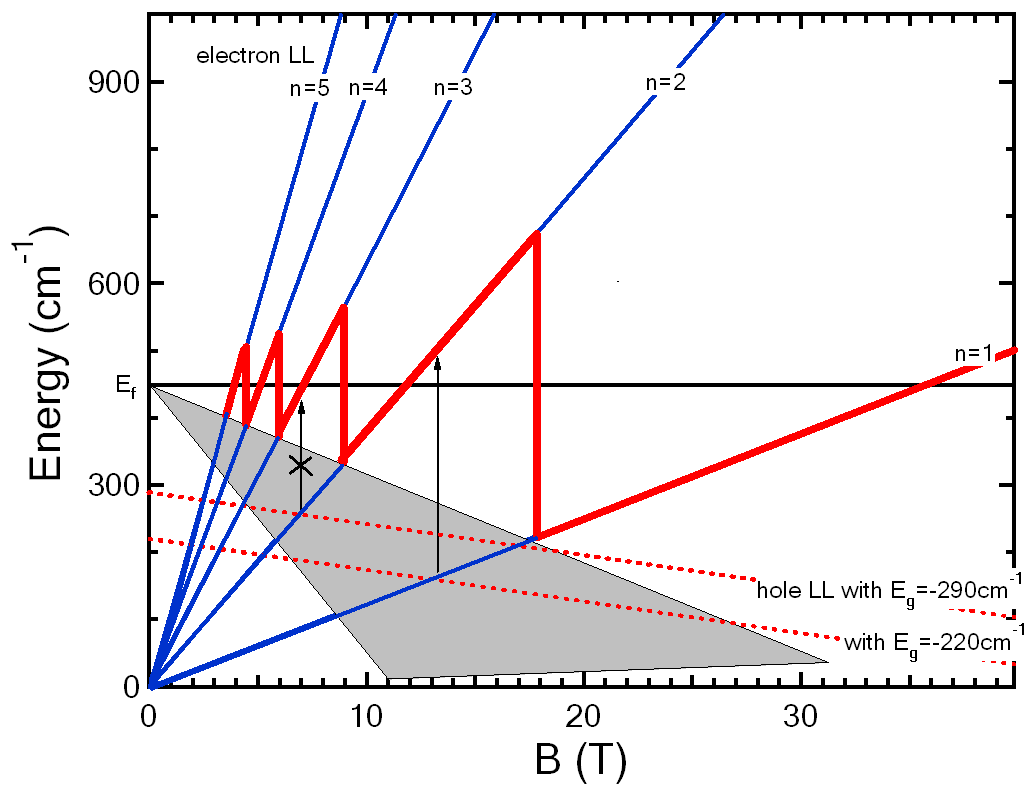}}}}\caption{Electron (blue solid line; $n=1-5)$ and
hole (red dotted line; $n=1$) Landau level structures vs magnetic field are
plotted using the following parameters: $m_{e}^{\ast}=0.037m_{e}$,
$m_{h}^{\ast}=0.1m_{e}$ and electron density $n_{e}=$ $8.6\times
10^{11}\operatorname{cm}^{-2}$. For convenience, energies are represented in
$\operatorname{cm}^{-1}$ and g-factors for both electron and hole levels are
ignored. The Fermi level oscillation is shown in a red thick solid line. The
shaded area defines the region where the next set of LLs are below the FS. Two
alternative values of the energy gap are discussed and shown in the two red
dotted lines. CR splitting has been observed around $B=14\operatorname{T}$,
and it coincides with the position of the antilevel crossing if $E_{g}%
=-220cm^{-1}$, thus shown with an upward arrow. CR splitting was not observed
around $B=7\operatorname{T}$, where an antilevel crossing is expected if
$E_{g}=-290cm^{-1}$, thus shown with the upward arrow crossed out. Higher hole
LLs are ignored in this plot.}%
\end{figure}

By selecting a small effective band gap $E_{g}$ ($\sim220%
\operatorname{cm}%
^{-1}$)\cite{Note1}, it can be clearly seen that any anomalies, due to either
models, will not result in multiple absorption modes at low magnetic fields,
since these anomalies are deep in the Fermi sea and the neighboring LLs are
occupied. This explains why no CR splitting was observed at low magnetic
fields in our samples or in the ones reported by Heitmann \textit{et
al}.\cite{Hei86} The electron densities in both cases are generally larger
than the ones used in other reports, resulting in a larger Fermi energy
$E_{f}$ as compared to $\left\vert E_{g}\right\vert $.

At intermediate magnetic fields, anomalies near the crossing point ($\sim14%
\operatorname{T}%
)$ can result in multiple absorptions through LL transitions between the $n=1$
and $n=2$ electron-like LLs, and from the hole-like LLs to the $n=2$
electron-like LLs, possibly responsible for the two broad modes observed
between $14%
\operatorname{T}%
$ to $18%
\operatorname{T}%
$. However, CR splitting at higher magnetic fields cannot be a result of
electron-hole LL mixing, since electron LLs will no longer be aligned to the
hole LLs at higher magnetic fields. This picture may explain why there exist
two distinct regimes for CR splitting. In the intermediate magnetic field
regime ($14%
\operatorname{T}%
$ to $18%
\operatorname{T}%
$), absorption modes are broad and weak with a larger energy separation, in
which the hybridization may have an effect on these modes; while in the high
magnetic field regime ($B\geqq20%
\operatorname{T}%
$), absorption modes are sharp and strong with a smaller energy separation, in
which hybridization has negligible effect.

Indeed, electron-hole LL mixing can occur at higher magnetic fields if the
hole levels are located at higher energies. (i.e. larger $\left\vert
E_{g}\right\vert $) A negative energy gap as large as $-90$m$%
\operatorname{eV}%
$ has been used to account for the multiple absorptions observed at high
magnetic fields.\cite{Pet04} Even by shifting the hole LLs up by less than
$10$m$%
\operatorname{eV}%
$, where the LL mixing can barely result in multiple absorptions at around $20%
\operatorname{T}%
$, LL mixing will also result in additional absorption at around $7%
\operatorname{T}%
$ through LL transitions between $n=2$ and $n=3$ LLs, which were not observed
in this series of samples. The position of the Fermi surface (FS) can place
the minigap deep in the Fermi sea, and naturally minimize the influence of the
hybridization even if the sample is strongly hybridized. In fact, most of the
reports in favor of the hybridization model have placed the FS between
electron and hole subbands, i.e. within the energy gap, which will definitely
result in multiple modes at very low magnetic
fields.\cite{Cla86,Sun92,Chi96,Yan97,Lak97,Mar99,Vas99,Pou99,Pet02,Pet04,Pet07}%

Another explanation for the lack of CR splitting at lower magnetic fields is
that the interposed barrier layer suppresses the strength of the
hybridization. No matter whether the FS is above the hole subbands or between
the electron and hole subbands, hybridization in these samples is too weak to
result in multiple modes even when the electron and hole LLs are aligned. The
latter appears to be a more viable explanation, since most of the band
calculations place the FS within the energy gap.

\subsection{Energy separation of the splitting}

The energy separation between the absorption modes appeared to be
field-independent when it was plotted over a wide range in previous reports.
Multiple absorption modes were obtained from a set of spin-split LLs in the
hybridization model. Although the magnetic-field induced spin splitting may
not be as significant as the antilevel crossing gap, the separation still
carries a Zeeman term which grows with increasing magnetic field.\cite{Chi96}
Upon taking a closer look at the data in the literature, the separation
between the modes is indeed increasing with increasing magnetic field defined
piecewise between the crossing points of the electron and hole
LLs.\cite{Pet04,Pet07} Since the region between two crossings expands with
increasing magnetic field (see Fig. 3), a magnetic-field dependent separation
will be more pronounced at high magnetic fields.

\begin{figure}[tb]
{{{\includegraphics[
natheight=11in,
natwidth=8.5in,
height=4in,
width=3.4in
]{C: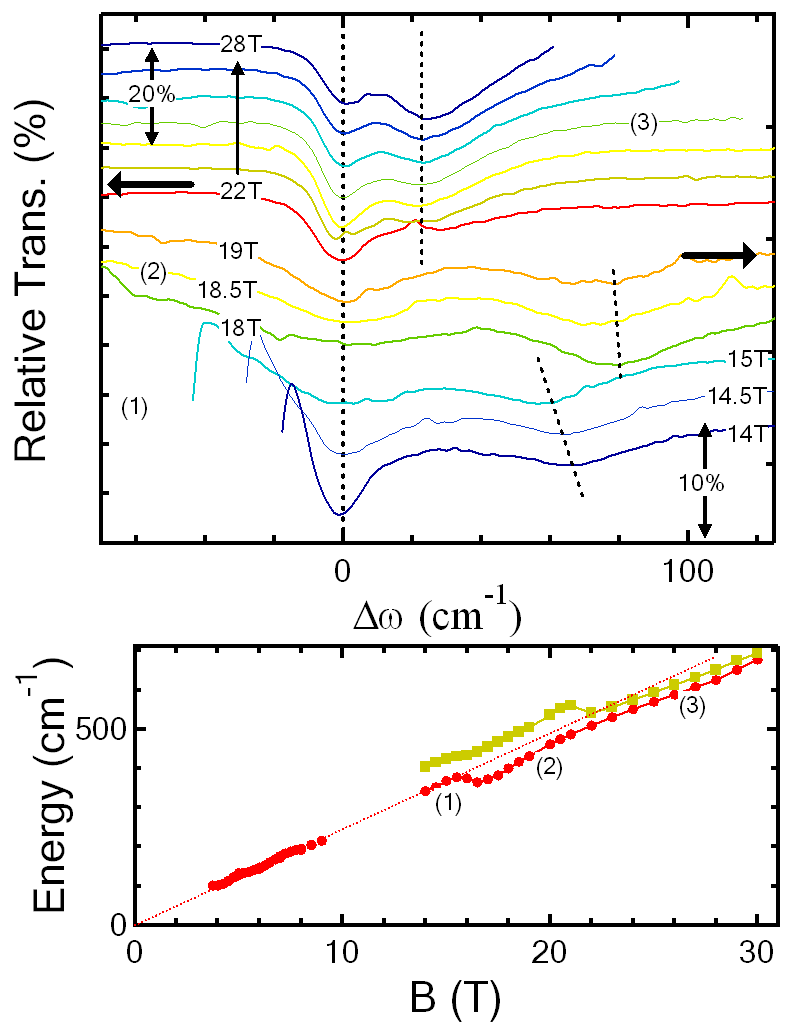}}}}\caption{Top: To demonstrate the evolution of the
energy separation between CRL and CRH more clearly, the spectra are
horizontally offset to have CRL aligned. The part near the restrahlen band and
the beamsplitter cutoff are removed for clarity. Bottom: The energies of the
absorption modes as a function of magnetic field are displayed. CRL and CRH
modes are represented in red circles and orange squares, respectively. The
energy separation is about $60\operatorname{cm}^{-1}$ in zone (1),
$80\operatorname{cm}^{-1}$ in zone (2) and $25\operatorname{cm}^{-1}$ in zone
(3). }%
\end{figure}

To demonstrate that the separation between the CR splitting, CRL and CRH, is
field-independent more precisely, the spectra were offset horizontally to
align CRL at $22%
\operatorname{T}%
$, as shown in Fig. 4 and the energies of the modes as a function of magnetic
field are plotted in the bottom. In fact, there exist three zones with three
distinct separation energies. In general, the energies of CRL and CRH increase
linearly with increasing magnetic field. Two abrupt drops in each trace divide
the region, where the CR splitting have been observed, into a succession of
three zones, labeled: zones (1), (2) and (3), respectively.\ At the lower end
of this region (zone 1), the separation energy is about $60%
\operatorname{cm}%
^{-1}$ before CRL makes an abrupt drop. The energy separation is then about
$80%
\operatorname{cm}%
^{-1}$ (zone 2) until CRH also makes an abrupt drop in energy. Above $22%
\operatorname{T}%
$, the separation is maintained at around $25%
\operatorname{cm}%
^{-1}$ (zone 3) as long as two modes are still observed up to $30%
\operatorname{T}%
$. These values are of the same magnitude as compared to the exciton binding
energies calculated by various works, but none have suggested a
field-dependent binding energy with discontinuities. The discontinuities in
mode energies have been observed in several
works.\cite{Che95,Kon97,Pet04,Pet07,Kon98} These two abrupt drops in energy
may be regarded as a part of the CR effective mass oscillation, and it will be
discussed in section E.

In zones (1) and (2), the energy separations between CRL and CRH are generally
larger than the separation in zone (3), which is inconsistent with the concept
that the CR splitting are a result of a pair of LLs with different spin
states. Moreover, within zones (1) and (2), the energy separation is again
slightly larger at lower fields, inconsistent with the nature of the Zeeman
splitting. In zone (3), the energy separation maintains a nearly perfect
field-independent energy separation up to $30%
\operatorname{T}%
$. At high magnetic fields, the influence of the spin splitting should be
discernibly visible. The lack of a magnetic-field dependent energy separation
in this work suggests that the CR splitting is not a consequence of a set of
spin-split LLs, and is thus unlikely to result from the hybridization of the
electron and hole wavefunctions. We here acknowledge that works in favor of
the exciton model have reported an energy separation, which was also slightly
larger at lower magnetic fields.\cite{Kon98}

\subsection{Effect of the thermal excitation}

Previous studies in these types of materials have revealed that multiple
absorptions will merge into one if the temperature is sufficiently
high.\cite{Che95,Kon97,Mar99,Pet07} In the exciton model, the higher-energy
transition was interpreted as an exciton internal transition, while the
thermal energy, overcoming the binding energy, breaks the bond between the
electron and hole in the exciton.\cite{Che95,Kon97} Marlow \textit{et al.}
then argued that the higher-energy transition did not decay fast enough for an
exciton with a binding energy about $12%
\operatorname{cm}%
^{-1}$; therefore, the decay of the higher-energy transition is a result of
populating the higher-energy spin-split LLs with increasing
temperature\cite{Mar99} and interpreted it in terms of the hybridization
model. Petchsingh \textit{et al.} then observed that multiple modes merged
into one at around $40%
\operatorname{K}%
$, while the sum of the absorption intensity increases with increasing
temperature.\cite{Pet07} The increase in the integrated intensity has been
interpreted to mean that the thermal excitations give access to transitions
associated with levels away from the minigap with lower degree of
hybridization.\cite{Pet07} In spite of how the temperature dependence was
interpreted in the past, multiple modes converge at high temperatures and
usually merged into one at even higher temperatures beyond the critical point
of modal convergence.

\begin{figure}[tb]
{{{\includegraphics[
natheight=8.5in,
natwidth=11in,
height=2.473in,
width=3.4in
]{C: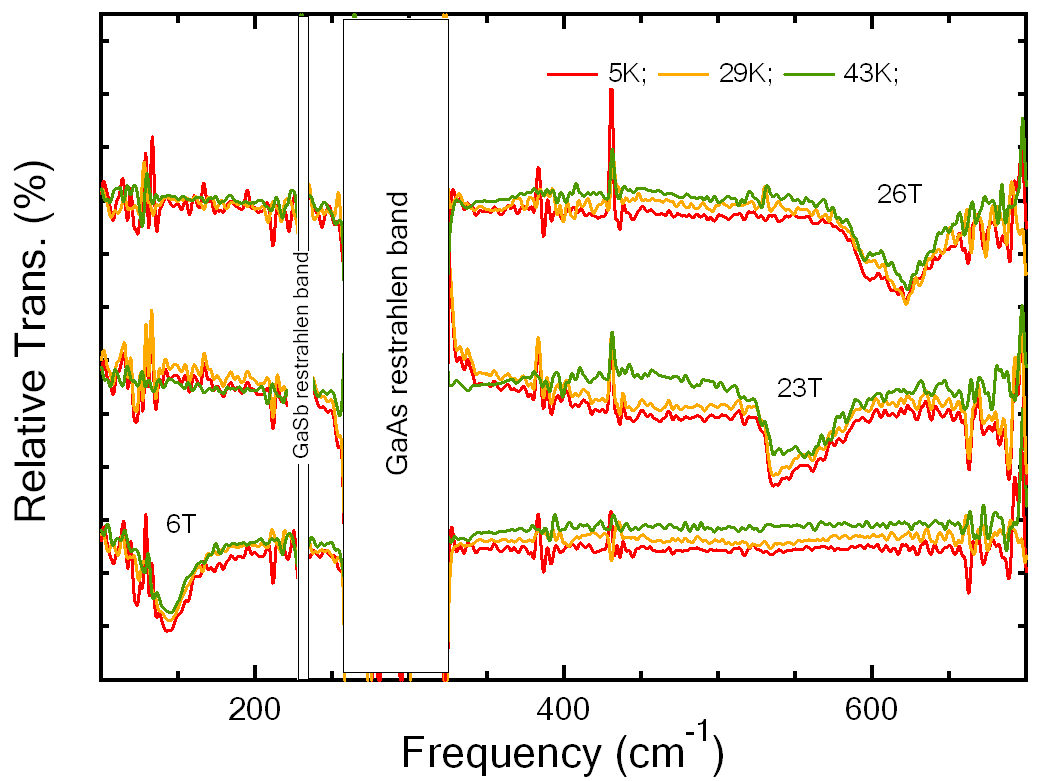}}}}\caption{The temperature-dependent magneto-FIR
spectra are displayed at three temperatures ($5$, $29$ and $43\operatorname{K}%
$) for three different magnetic fields. ($6$, $23$ and $26\operatorname{T}$)
The spectra of different temperatures are slightly offset to make each trace
more resolvable.}%
\end{figure}

A set of spectra measured at $6$, $23$ and $26%
\operatorname{T}%
$ for three different temperatures are displayed in Fig. 5 to demonstrate the
temperature dependence of the CR splitting. One can instantly see that the
absorption modes are insensitive to increasing temperature. These modes are
rather robust against rising thermal excitations, which suggests that both
modes result from free-carrier excitations with the FS far away from the
minigap, consistent with the picture illustrated in Sec. A.

As indicated in Sec. A, the minigap, if existed, is deep in the Fermi sea and
the states around minigaps are occupied. Since the FS is pinned at the
corresponding electron LLs with a LL separation around several hundred $%
\operatorname{cm}%
^{-1}$, thermal excitation at $40%
\operatorname{K}%
$ can hardly distort the FS at high magnetic fields. Though it is possible
that the minigap lies closer to the FS at intermediate fields ($14$-$18%
\operatorname{T}%
$), the spectra at these fields are not displayed due to the reduced
signal-to-noise ratio at higher temperatures. The broad peaks observed at
intermediate fields are close to the noise level, but the results are more
likely to be temperature-independent.

\subsection{Effect of parallel field}

The nature of the hybridization model is an antilevel crossing gap as a result
of Pauli exclusion principle, creating minigaps near the resonances of
electron and hole LL wavefunctions. A parallel magnetic field can cause a
phase shift between electrons and holes, moving the electron and hole
dispersion relations in opposite directions in $k$%
-space.\cite{And82,Hei93,Pou99} This shift decouples the dispersion relations
and eliminates the formation of the minigap. This phenomenon was
experimentally confirmed in the InAs/GaSb CQWs,\cite{Lak97,Mar99,Pou99,Pet07}
and was considered as plausible evidence in support of the formation of
minigaps in these types of materials. Similar to what happened to the
absorption modes when the temperature is increased, multiple absorption modes
merge into one with increasing parallel magnetic
field;\cite{Yan97,Mar99,Pou99,Pet07} while Petchsingh \textit{et al.} reported
an increase of the integrated intensity occurred simultaneously\cite{Pet07}
not unlike the one discussed in the previous section. A parallel magnetic
field as small as $7%
\operatorname{T}%
$ suppresses the hybridization and destroys the formation of the minigaps even
in the strongly hybridized samples.\cite{Mar99}

\begin{figure}[tb]
{{{\includegraphics[
natheight=11in,
natwidth=8.5in,
height=4in,
width=3.4in
]{C: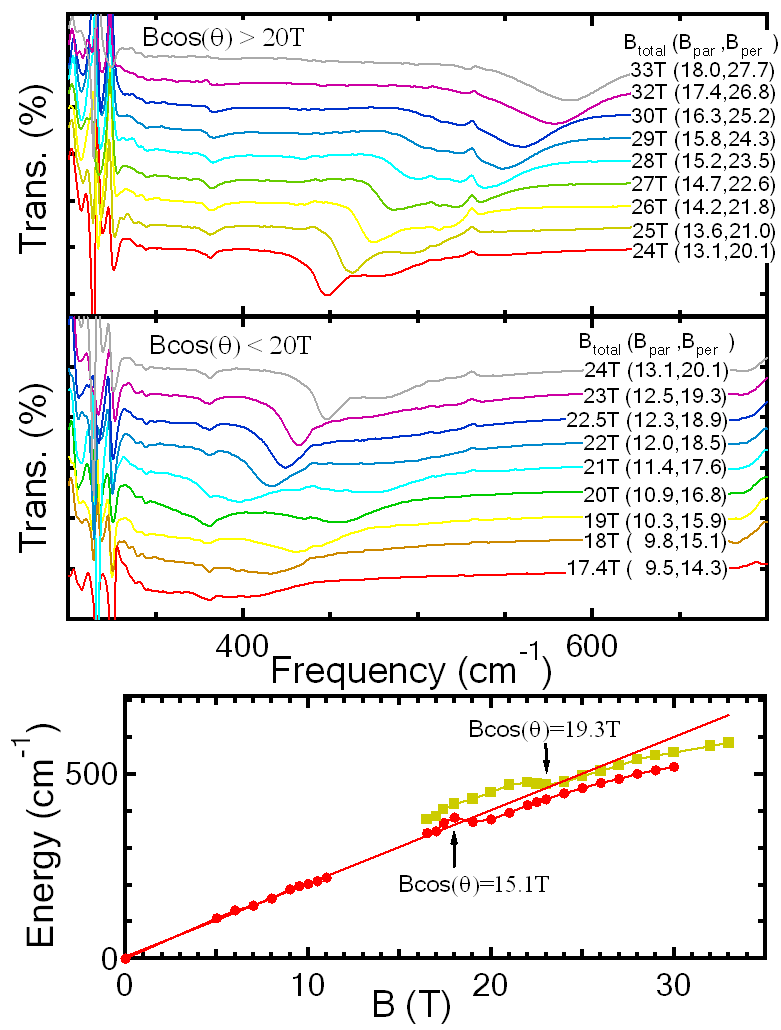}}}}\caption{Top: Magneto-FIR spectra measured in
tilted geometry ($33^{\text{o}}$) from $17.4\operatorname{T}$ to
$33\operatorname{T}$ are displayed. The spectra are divided into two groups
based on the strength of the perpendicular magnetic field. The magnitudes of
the parallel ($B_{par}$) and perpendicular ($B_{per}$) fields at a given total
field ($B_{total}$) are shown by the values in the bracket. Bottom: The
energies of the absorption modes as a function of magnetic field are
displayed. CRL modes are shown in red circles and CRH modes are shown in
orange squares. The energy separation is about $40\operatorname{cm}^{-1}$ in
zone (1), $75\operatorname{cm}^{-1}$ in zone (2) and $36\operatorname{cm}%
^{-1}$ in zone (3).}%
\end{figure}

The sample is placed on a wedge with the normal direction of the sample tilted
at an angle from the direction of the magnetic field. The tilt angle was
determined to be $33%
\operatorname{{{}^\circ}}%
$ by fitting the CR energies at low magnetic fields. A set of magneto-FIR
spectra measured in the tilted geometry is displayed in Fig. 6 and the
energies of the CRL and CRH modes as a function of the total magnetic field
are displayed in the bottom. The magnitudes of the parallel ($B_{par}$) and
perpendicular ($B_{per}$) fields are shown in the bracket and a parallel
magnetic field as large as $18%
\operatorname{T}%
$ has been reached. The spectra were divided into two groups; the spectra with
a perpendicular field larger than $20%
\operatorname{T}%
$ are displayed in the top graph while the ones below $20%
\operatorname{T}%
$ are displayed in the bottom. Twenty tesla was the magnitude of the magnetic
field which divided the CR splitting into a sharp-peak region and a broad-peak
region in the Faraday geometry.

As compared to the spectra measured in the Faraday geometry, the absorption
modes become sharp and strong at around the same total magnetic field, instead
of the same perpendicular field, suggesting that the sharp-peak behavior, or
the narrowing of the linewidth, are induced by the total field, independent of
the tilt angles. However, the abrupt drops of mode energies occur at around
the same perpendicular fields; in other words, they occur at the same filling
factors or the same LL separations.

The CR splitting are observed at a parallel magnetic field higher than $10%
\operatorname{T}%
$, while it was shown to be sufficient to suppress the formation of the
minigaps in strongly hybridized samples.\cite{Pou99,Pet07} By comparing the
spectra of similar perpendicular fields, but different parallel fields, CRH in
comparison to CRL becomes slightly stronger, which can simply be a result that
their relative strength are mainly determined by the total field, instead of
the perpendicular field.

The energies of CRL and CRH generally scale with tilting angle $\theta$ as
$\cos\theta$ with an additional suppression at high fields, possibly due to
the additional confinement introduced by the high parallel fields. Additional
confinement reduces the effective well width, thus resulting in a larger
effective mass; in other words, a lower CR energy. The energy separation is
about $40\operatorname{cm}^{-1}$ in zone (1), $75\operatorname{cm}^{-1}$ in
zone (2) and $36\operatorname{cm}^{-1}$ in zone (3). The first two are
slightly smaller than their counterparts in the Faraday geometry, while the
last one is slightly larger than its counterpart. The energy separation is
slightly smaller at intermediate magnetic fields, which may be due to the
closing of the minigaps by increasing parallel magnetic field; however, it
also demonstrates that the effect of the hybridization is secondary to the
major factor causing the CR splitting. At high magnetic fields, the separation
becomes larger, which cannot be explained by the hybridization model, or the
concept of spin-split electron CR.

In the tilted geometry, parallel magnetic field can also cause the LLs of
different electron subbands to couple, known as resonant subband LL
coupling.\cite{Sch83} A quick estimate places the electron's 2nd subband
$105$m$%
\operatorname{eV}%
$ above the first, which is too high to produce an effect in measurements
below $33%
\operatorname{T}%
$.

\subsection{Oscillation of CR amplitude, width and effective mass}

The oscillation of CR was found to be more pronounced in the InAs-based
2DS\cite{Hei86,Yan93,Scr93,Kon94,Kon97} than the one in the GaAs-based
systems.\cite{Eng83} On one hand, CR linewidth maxima were observed at even
filling factor $\nu$'s and the minima at odd $\nu$'s,\cite{Hei86,Kon94,Kon97}
attributed to the $\nu$-dependent screening of impurity scattering by the 2D
electron gas.\cite{And75,Sar81,Las83} On the contrary, linewidth maxima were
observed for odd $\nu$'s in the spin-resolved CR
measurements,\cite{Yan93,Scr93} attributed to the nonparabolicity
effect.\cite{Han88} The driving mechanism behind such a pronounced CR
oscillation particularly in the InAs-based 2DS is not fully understood;
nonetheless, CR oscillation can still be used to reveal principal parameters,
such as carrier densities, which in this case, help to determine which mode,
CRL or CRH, is more likely to result from the typical electron CR.

Unlike the multiple absorption modes reported in the literature, the observed
CR splitting is insensitive to increasing temperature or parallel magnetic
fields. As a result, both of them can be a conventional electron CR. To
understand the origin of the CR splitting, it is very important to identify
which one results from the typical electron CR. An electron-hole coupling is
likely to have an attractive interaction, which tends to result in an
absorption at higher energies as more energy is needed to break the coupling;
while an electron-electron coupling is likely to have a repulsive interaction,
leading to an absorption mode at lower energies. \begin{figure}[pt]
{{{{{\includegraphics[
natheight=11in,
natwidth=8.5in,
height=4in,
width=3.4in
]{C: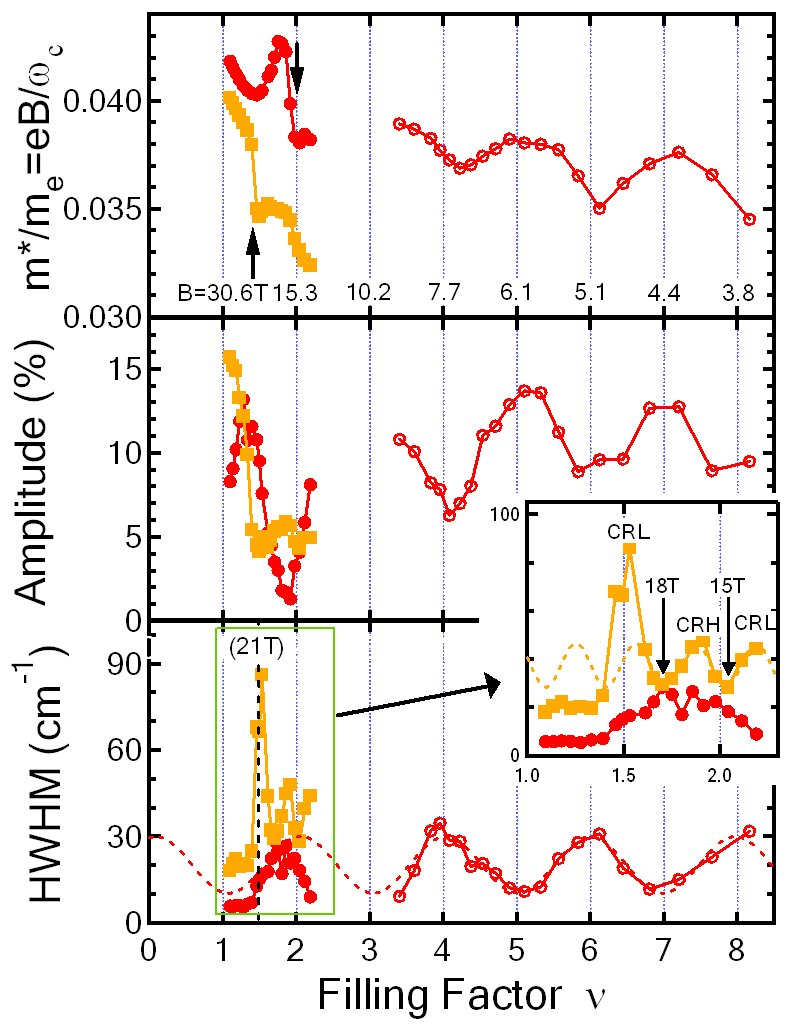}}}}}}\caption{The oscillation of effective mass,
amplitude and linewidth for CRL (red open circles for lower fields, where only
one absorption mode was observed and red filled circles for higher fields) and
CRH (orange filled squares) as a function of filling factor is plotted. The
sudden rises in the CR effective mass for CRL and CRH modes are indicated by a
downward and an upward arrow, respectively. From the CR oscillation, the
electron density was determined to be $7.4\times10^{11}\operatorname{cm}^{-2}%
$. Inset: Expanded view of the linewidth oscillation at high magnetic fields.}%
\end{figure}

The oscillations of CR effective mass, amplitude and linewidth (half width at
half maximum) are displayed in Fig. 7 and the period of the oscillation agrees
well with the electron density of $7.4\times10^{11}\operatorname{cm}^{-2}$
with linewidth maxima for even-valued filling factor $\nu$'s. By using the CR
effective mass oscillation observed at lower magnetic fields, where only a
single absorption has been observed, one can extend the oscillations to higher
magnetic fields to determine which one is more likely to be electron CR. CRL
has an oscillation pattern consistent with the ones found at low magnetic
fields; therefore, we intend to believe that CRL results from the typical
electron CR in the InAs layer, while CRH is an induced mode. Since it has
produced an additional absorption mode at a higher energy, an attractive
interaction is more likely to be the cause. Electron-hole interaction is a
likely cause, while the hybridization model results from an antilevel
crossing, which comes from the repulsive interactions between the states. In
the works supporting the hybridization model, a new mode emerges from the
lower energy side,\cite{Pet04} which illustrates that the interaction is repulsive.

The CRL's effective mass oscillates with filling factors and a sudden rise in
effective mass occurs at around $\nu\sim2$, corresponding to the abrupt drop
in CRL energy in Fig. 4; then, it is possibly approaching another effective
mass maximum near $\nu\sim1$. The sudden rise in effective mass may be related
to the unique properties of the $\nu=2$ quantum Hall states, which has been a
subject of interest recently.\cite{Pia99} Though the region where the CRH mode
has been observed is not wide enough to exhibit significant periodic behavior,
a similar effective mass jump has been observed at around $21%
\operatorname{T}%
$, corresponding to the abrupt drop in CRH energy. It has been shown that both
of these jumps are either filling factor or perpendicular field dependent in
Sec. III. D. One can hardly correlate $\nu=1.5$ and $\nu=2$ to one another and
to a common physical origin. It is reasonable to assume that the effective
mass jump of CRH mode also occurs at around $\nu=2$, but from a region of
slightly higher carrier density. The implication of this concept will be
discussed in Sec. IV C.

The amplitude oscillation at high magnetic fields, i.e. lower filling factors,
is more driven by the exchange of the integrated intensity between the CR
splitting, instead of the $\nu$-dependent oscillations. CRL's amplitude
decreases while CRH's increases between $14%
\operatorname{T}%
<B<17%
\operatorname{T}%
$, then the process is reversed between $17%
\operatorname{T}%
<B<21%
\operatorname{T}%
$. Both amplitudes increase with increasing magnetic field between $21%
\operatorname{T}%
<B<23%
\operatorname{T}%
$, corresponding to an anomalous linewidth narrowing at $21%
\operatorname{T}%
$. (transition from the broad-peak to the sharp-peak regions) In the end,
CRH's amplitude continues to increase for $B\geqslant23%
\operatorname{T}%
$, while CRL's amplitude decreases with increasing magnetic fields. The
evolution of the CR amplitudes with increasing magnetic fields tends to mimic
the evolutionary profile of the absorption strength observed in Fig. 2.

The linewidth oscillation of CRL and CRH at high magnetic fields are expanded
in the inset, which display an interesting pattern. While CRL's linewidth
generally oscillates with the filling factors at high magnetic fields, CRH's
oscillates at a much faster rate. If this rate is taken seriously, such a fast
oscillation corresponds to an electron density as large as $4.75\times10^{12}%
\operatorname{cm}%
^{-2}$ with linewidth maxima at even filling factors ($10$, $12$, and $14$)
and minima at odd filling factors ($9$, $11$, and $13$), opposite to the CRL's
linewidth oscillations. It is very unlikely that such a dense region (by an
order) is formed in this system. The maxima of the CRH's linewidth corresponds
to the fields, where one of the two modes is dominating, while the two minima
correspond to the fields where the two modes are of nearly equal strength. It
suggests that the linewidth oscillation for CRH mode is also strongly affected
by the intensity exchange between the CR splitting.

At around $21%
\operatorname{T}%
$, linewidths of both modes collapse, though it is less significant for CRL,
since its linewidth is decreasing from $\nu=2$ to $\nu=1$. At very high fields
($B\geq23%
\operatorname{T}%
$), both linewidths remain small and become independent of the magnetic fields
or filling factors. The sharp reduction of photoluminescence linewidth was
attributed to a phase transition of the exciton system into an ordered
state.\cite{Fuk90} This sharp reduction of CR linewidth could be an effect
parallel to that found by Fukazawa \textit{et al.}, but we lack the evidence
supporting that a critical temperature $T_{C}$ is associated with the
narrowing of the CR linewidth. The CR splitting appear to be
temperature-independent up to $43%
\operatorname{K}%
$ and it is the highest temperature attainable with our present high-field probe.

\section{Discussion}

Many aspects of this work have been discussed and compared with the two most
widely used models, the exciton model and the hybridization model. As
indicated in the previous sections, neither one can account for the features
found in this paper. Though they appear to be quite different, interlayer
Coulomb interaction and hybridization of electron-hole wavefunctions share
many similar properties. By considering the electron-hole system's Hamiltonian
as a simple $2\times2$ matrix as illustrated by Petchsingh \textit{et
al.},\cite{Pet04} hybridization effect and interlayer Coulomb interaction will
both appear in the off-diagonal term, thus resulting in similar types of
energy characteristics, though an interlayer Coulomb interaction is possibly
more complicated. The hybridization model is realized as two intercepting
dispersion relations; one belongs to the electron and the other to the holes.
The minigaps are formed at the intercepts of the electron and hole dispersion
relations, i.e. where the $k$ wavevectors belonging to different dispersion
relations are equal. Coulomb interactions, in the form of $1/k^{2}$, should
also have a prominent effect at the intercepts. When the hybridization model
is realized in an energy diagram shown in Fig. 3, a minigap of several m$%
\operatorname{eV}%
$ is formed when the electron and hole LLs are aligned with increasing
magnetic field. By contrast, an exciton is formed with a characteristic energy
of several meV when the single particle energy gap becomes smaller than the
binding energy, i.e. when the electron and hole LLs are nearly aligned. For
the hybridization model, the $\mathbf{k}\cdot\mathbf{p}$ model has established
an energy diagram for the multiple absorption modes that agree well with the
observed mode energies, which may also reflect what is going to occur if the
interlayer Coulomb interaction is included in the $\mathbf{k}\cdot\mathbf{p}$ model.

\subsection{A Modified Exciton Model}

We would like to suggest a small tweak to the original exciton model. As
indicated in the Sec. III E, the CRL mode is likely to result from the
conventional electron CR, which is a LL transition from $n=0$ to $n=1$.
Instead of having exciton internal transitions, CRH is an exciton
pair-breaking excitation. The exciton absorbs the photon energy, breaking the
electron-hole pair and the electron makes a LL transition from $n=0$ to $n=1$
LL. As a result, its magnetic-field dependence is similar to the electron-LL
transition, but is instead offset by the binding energy. Since the spin-split
ground state LL is fully occupied up to $30%
\operatorname{T}%
$, a pair broken by the thermal excitations cannot place the electron back to
the ground state, thus rendering it forbidden. Extra thermal energy is needed
in order to break the pair and place the electron in the next available LL,
which is separated by about $200%
\operatorname{cm}%
^{-1}$.($25%
\operatorname{T}%
$ with $g\sim-8$) As a result, both modes are insensitive to the temperature
changes up to $43%
\operatorname{K}%
$ within the magnetic field range investigated. Unfortunately, this concept
suffers a severe setback; since the hole density is much smaller than the
electron density. Even if each hole is bound to an electron, the CRH mode
should remain the weaker mode of the two; which contradicts observation -
unless, we assume that a hole is bound with several electrons forming the
many-particle state, which release an electron after it absorbs the photon energy.

\subsection{Spin-Split CR}

As indicated in Sec. III A, the hybridization model cannot account for the CR
splitting observed at high magnetic fields, or it will result in multiple
absorption mode at low fields. It does not seem to rule out the possibility
that the CR splitting are caused by the spin-split LLs as a result of a
difference in the $g$-factors of LLs with different LL index; in this case,
$n=1$ and $n=2$ LLs. We have basically ruled out the spin-split CR as a
possible cause for the CR splitting when we ruled out the hybridization model;
since it also arose from a set of spin-split LLs. At this juncture, we will
discuss several additional discrepancies between the observed features and the
features that is expected if the CR splitting are a result of the spin-split LLs.

In the single particle picture, such an effect is basically the Zeeman
splitting, which increases linearly with increasing magnetic field as $\Delta
g\cdot\mu_{B}\cdot B$, where $\Delta g$ is the difference in the $g$-factor
and $\mu_{B}$ the Bohr magneton. The observed energy separation is
field-independent at high magnetic fields, while being slightly larger at
lower magnetic fields. To account for the observed energy separation, $\Delta
g$ needs to be rather complicated, which has to be field-dependent as $1/B$ at
high fields, and as $1/B^{\gamma}$ with $\gamma>1$ piecewise defined at the
intermediate fields. Additionally, this must occur between $\nu\gtrsim2$ and
$v=1$.

The lineshapes of the CR splitting are not symmetrical, since CRH always has a
larger linewidth. This would require the system to have a
spin-direction-selective scattering mechanism. The strength of the two modes
are nearly equal for three different fields, which are $15.5,$ $19$ and $25%
\operatorname{T}%
$, respectively. The spin-split CR should have nearly equal contributions at
$\nu\sim2$ and then one of them, usually the lower-energy transition,
increases at the expense of the other one with increasing magnetic field. The
intensity of the two modes of the CR splitting exchange several times between
$\nu=2$ and $\nu=1$ and there is no way to account for all three magnetic
fields, where the intensity of the two modes are nearly equal to each other.
The CR splitting should start to appear at around $\nu=4$, which is around $8%
\operatorname{T}%
$ with at least three different transition energies. One is at around the
typical CR energy if the $n=2$ and $n=3$ LLs have the same $g$-factor and a
pair of modes offset by $\pm\Delta g\cdot\mu_{B}\cdot B$. Such splitting has
not been observed in this system. In addition, the features observed when the
sample is tilted are also against the spin split CR; in particular, the energy
separation increases with increasing tilt angle $\theta$, while it should be
tilt-angle-independent for spin splitting.

\subsection{Spontaneous Phase Separation}

It has been proposed that electrons were trapped by the localized states on
the surface\cite{Mik75,Kot75,Che90} or in puddles between repulsive
scatterers,\cite{Ric89} forming regions of higher electron densities, in which
the CR energy of the weakly bound electrons in the harmonic traps is shifted
to a value $\omega^{2}=\omega_{c}^{2}+\Omega^{2},$ where $\Omega$ is related
to the curvature of the confining potential. At sufficiently high magnetic
fields, both of the CR and the confinement shifted CR increase linearly with
increasing magnetic field, as if the latter offsets from the the former by a
constant energy separation. Moreover, the energy separation between the two,
$\omega_{c}$ and $\omega$, will be larger at lower magnetic fields, but nearly
a constant at high magnetic fields.

\begin{figure}[tb]
{{{\includegraphics[
natheight=11in,
natwidth=8.5in,
height=4in,
width=3.4in
]{C: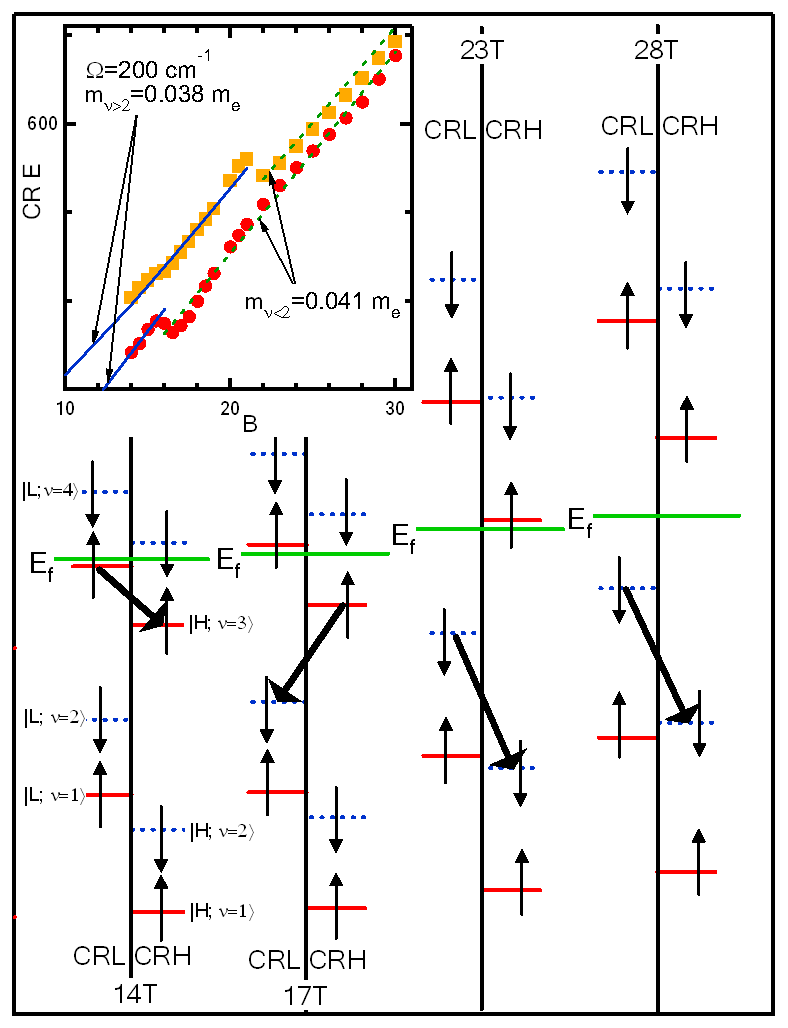}}}}\caption{Energy diagram of the LLs in the puddles
(CRH) and in the envirmoent. (CRL) The energies of the CRH LLs are shifted by
about -$200\operatorname{cm}^{-1}$ to account for an energy trap for the
electrons, resulting in the difference in the electron densities. The thick
vertical lines represent the interfaces between them and the energies of the
LL are plotted to scale based on the measured energies of CRL and CRH modes.
To distinguish the spin up (red solid line) and spin down (blue dashed line)
LLs, a $g$-factor of $-8$ is used to calculate the Zeeman splitting. The
positions of the FS at different fields are represented in greed solid lines
and the bold arrows show the direction of the electron transfer. Inset: The
energies of the CRL and CRH modes are fitted by a phenomenological model, in
which the CR oscillation is ignored and the effective mass jump near $\nu=2$
is approximated by a step function. When $\nu>2$ (blue solid lines), the
electron effective mass is $0.038m_{e}$ and $0.041m_{e}$ when $\nu\leq2$.
(green dotted lines) With $\Omega=200\operatorname{cm}^{-1}$, the energies of
CRL and CRH mode can be fitted by the magnetic field dependence of the
electron CR and the confinement shifted CR, respectively. }%
\end{figure}

It has been argued previously (III E) that the sudden rises in the CR
effective mass of both modes may result from a common cause. While it is known
that CRL's effective mass jump occurs when the filling factor approaches
$\nu=2$, the jump of CRH mode may result from regions of higher electron
density, which approaches $\nu=2$ at higher fields. We propose that a
spontaneous phase separation is induced by increasing magnetic field and the
2DES splits into two phases: puddles with the electron density of around
$10.2\times10^{11}\operatorname{cm}^{-2}$ surrounded by the region with the
electron density of around $7.4\times10^{11}\operatorname{cm}^{-2}$.
Additional confinement shifts up the CR energy in the puddles and the edge
scattering of the puddles increases the overall CRH linewidth.

Although it appears that the confinement shifted CR is described by a simple
guadratic sum, it is difficult to formulate the CR energy itself due to strong
CR oscillation and nonparabolicity in InAs 2DES. Disregarding CR oscillation
and nonparabolicity at high fields, we describe the CR energy
phenomenologically by approximating the effective mass jump near $\nu=2$ with
a step function. CR effective mass is $0.038m_{e}$ when $\nu>2$ and
$0.041m_{e}$ when $\nu\leq2$. The energy of CRH mode can then be described by
the confinement shifted CR at an appropriate filling factor for puddle's
carrier density is higher. CRL and CRH mode can be described fairly well by
this simple phenomenological model, as shown in the inset of Fig. 8 with
$\Omega=200%
\operatorname{cm}%
^{-1}$.

Most importantly, this concept can also account for the exchange of the
integrated intensity from $14%
\operatorname{T}%
$ ($\nu\gtrsim2$) to $30%
\operatorname{T}%
$ ($\nu\sim1$), which is difficult to explain in terms of the Fermi level
oscillation. An energy diagram for the LLs in the puddles (CRH) and in the
environment (CRL) is established in Fig. 8 to demonstrate the transfer of the
electrons, which leads to the exchange of intensity. The energies of the LL
are plotted to scale based on the measured energies of CRL and CRH modes at
different magnetic fields. A $g$-factor of $-8$ is applied to the calculated
LLs,\cite{Nil06} in order to distinguish LLs of different spins states. The
LLs in puddles (CRH) are offset by -$200\operatorname{cm}^{-1}$ to represent
an energy trap for the electrons, resulting in the electron density
difference. With increasing magnetic field, LLs in the puddles and in the
environment pass through the FS at different magnetic fields, resulting in the
exchange of the integrated intensity between the CR splitting.

To efficiently and clearly label the LLs in the puddles and in the
environment, each LL will be labeled as $\left\vert H\text{ or }L\text{; }%
\nu=?\right\rangle $, in which the first part labels whether it is a LL in the
puddles or in the environment. The second part labels the LL by the filling
factor, instead of the Landau level index for convenience.

At around $14%
\operatorname{T}%
$, $\left\vert H\text{; }\nu=4\right\rangle $ passes the FS, and electrons
fill up to $\left\vert L\text{; }\nu=3\right\rangle $. The electrons transfer
from $\left\vert L\text{; }\nu=3\right\rangle $ to $\left\vert H\text{; }%
\nu=3\right\rangle $, when empty states become available with increasing
magnetic fields, since $\left\vert L\text{; }\nu=3\right\rangle $ is higher in
energy than $\left\vert H\text{; }\nu=3\right\rangle $; therefore, the
intensity of CRH increases while the intensity of CRL decreases.

At around $17%
\operatorname{T}%
$, the transfer of the integrated intensity is reversed, since $\left\vert
L\text{; }\nu=3\right\rangle $ passes the FS. The electrons now transfer from
$\left\vert H\text{; }\nu=3\right\rangle $ to $\left\vert L\text{; }%
\nu=2\right\rangle $, since $\left\vert L\text{; }\nu=2\right\rangle $ is
lower in energy than $\left\vert H\text{; }\nu=3\right\rangle $; therefore,
the intensity of CRH increases while the intensity of CRL decreases.

Likewise, the transfer of the integrated intensity is reversed again at around
$23%
\operatorname{T}%
$ when $\left\vert H\text{; }\nu=3\right\rangle $ passes the FS and the
electrons transfer from the $\left\vert L\text{; }\nu=2\right\rangle $ to
$\left\vert H\text{; }\nu=2\right\rangle $. The electrons will continue to
pool into the puddles until $\left\vert L\text{; }\nu=2\right\rangle $ passes
the FS.

The size of the puddles can be estimated based on the equation proposed by
Mikeska \textit{et al.}.\cite{Mik75} If the depth of the trap is around
$200\operatorname{cm}^{-1}$, the average diameter of the puddles is about $250%
\operatorname{\text{\AA}}%
$\textbf{.} The temperature dependence of the CR splitting can be easily
explained if the triple point of the phase separation is much higher than $40%
\operatorname{K}%
$. A possible mechanism involves the holes in the GaSb layer, which trap the
electrons into puddles through the interlayer Coulomb interaction. A quick
estimate for the depth of the electron trap yields a value in the appropriate
range provided the band bending and the in-plane screening are ignored. Kallin
and Halperin\cite{Kal85} had previously discussed an impurity mode induced by
the impurity scattering; while, in this case, the impurities are the holes on
the other side of the barrier.

\section{Summary}

InAs/GaSb based type II and broken-gap quantum wells have been extensively
studied in the past as a promising candidate for finding BEC of excitons in
bilayer 2DS without photoexcitation. The works in the past have revealed
multiple absorption modes near CR which were interpreted in two different
models. On one hand, these modes indicated the formation of stable exciton
phase, which exhibits an additional mode at the exciton's $1s-2p$ internal
transitions. On the other hand, these modes seem to result from the formation
of the minigaps due the hybridization of the electron and hole wavefunctions.
Recent studies in these types of materials have been strongly in favor of the
hybridization model.

We have studied a series of weakly hybridized InAs/AlSb/GaSb CQWs, in which
the hybridization effect is minimal by having narrower well width and an AlSb
barrier interposed between InAs and GaSb layers. We have observed a pair of
modes, only at magnetic fields higher than $14%
\operatorname{T}%
$, in this series of samples; whereas, only a single absorption mode,
attributed to electron CR, was found in the weakly hybridized samples in the past.

The CR splitting exhibit several features, which are very different from those
reported in the past. The energy separation is slightly larger at lower
magnetic field, while it is magnetic-field independent at high magnetic field.
The CR splitting were robust against increasing thermal excitations and
increasing parallel magnetic fields. These key differences are inconsistent
with the two models proposed in the past.

Several ideas have been proposed and discussed in this paper. The concept, of
spontaneous phase separation, seems to account for most of the features
observed in this work. The phase separation can be \textbf{a} result of an
electron trap: a consequence of the interlayer Coulomb interaction from the
holes in the GaSb layer, which shows that the electron-hole interaction played
a vital role in this system.

\begin{acknowledgments}
This research was supported by contract
$\backslash$%
\# FA 9453-07-C-0207 of AFRL. The measurement was performed at NHMFL at
Tallahassee, supported by the National Science Foundation and the state of
Florida. One of us (W.X.) was supported by the Chinese Academy of Sciences and
National Natural Science Foundation of China.
\end{acknowledgments}

\end{document}